\documentstyle[prd,epsfig,aps,preprint]{revtex}
\begin{document}

\draft
\title{On the existence of turning points in D-dimensionsal Schwarzschild-de Sitter and
anti-de Sitter spacetimes}

   \author{C. H. G. Bessa$^1$\footnote{chbessa@dfte.ufrn.br} and
   J. A. S. Lima$^{1,2}$\footnote{limajas@dfte.ufrn.br,
   limajas@astro.iag.usp.br}}

  \smallskip
\address{~\\ $^1$Departamento de F\'{\i}sica Te\'orica e Experimental, \\Universidade Federal
             do Rio Grande do Norte, C.P. 1641, 59072-970, Natal,
             RN, Brasil
             $^2$Instituto de Astronomia, Geof\'{\i}sica e
Ci\^encias Atmosf\'ericas, Universidade de S\~ao Paulo\\ Rua do
Mat\~ao, 1226 - Cid. Universit\'aria, 05508-900, S\~ao Paulo, SP,
Brasil\\
                  }

\date{\today}
\maketitle

\vskip 1.5cm

\begin{abstract}
We investigate the motion of a test particle in a d-dimensional,
spherically symmetric and static space-time supported by a mass
$M$ plus a $\Lambda$-term. The motion is strongly dependent on the
sign of $\Lambda$. In Schwarzschild-de Sitter (SdS) space-time
($\Lambda > 0$), besides the physical singularity at $r=0$ there
are cases with two horizons and two turning points, one horizon
and one turning point and the complete absence of horizon and
turning points. For Schwarzschild-Anti de Sitter (SAdS) space-time
($\Lambda < 0$) the horizon coordinate is associated to a unique
turning point.
\end{abstract}


\newpage
\section{Introduction}

The best theory for describing gravitational interaction is the
classical general relativity, and the prediction of black holes is
usually considered one of its major triumphs. Many observational
aspects of black holes concern the properties of time-like and
null geodesics. In particular, the study of time-like geodesics in
spherically symmetric and static spacetimes (Schwarzschild and
Reissner-Nordtr\"{o}m (RN) solutions) has been discussed by
several authors\cite{jhm,lr}. Such studies lead to a reasonable
comprehension about the motion of a test particle near or inside
the horizon of uncharged (charged) black holes.

On the other hand, Einstein introduced the $\Lambda$-term in 1917,
and soon after de Sitter found a well behaved vacuum static
spherically symmetric solution for the modified equations. The
$\Lambda$-term alters considerably the solutions of the field
equations. In Schwarzschild and RN solutions, for example, the
resulting spacetimes have an extra event horizon. More recently,
in the cosmological context, the $\Lambda$-term has been
interpreted as a net vacuum energy density of all the existing
quantum fields, and, as suggested by the recent observations, may
be responsible by the present accelerating stage of the
Universe\cite{Perl}.

In this work we analyze some properties of time-like geodesics in
spherically static spacetimes. More precisely, we discuss the
existence of turning points in the Schwarzschild spacetime with
positive and negative $\Lambda$. Such spacetimes are usually
referred to as Schwarzschild-de Sitter (SdS) and
Schwarzschild-anti de Sitter (SAdS), and for the sake of
generality we consider the d-dimensional extension of both cases.

The geometry of a d-dimensional spherically symmetric and static
spacetime with a $\Lambda$-term reads

\begin{equation}
\label{met1} ds^{2} = f(r)c^{2}dt^{2} - f^{-1}(r)dr^{2} -
r^{2}d\Omega^{2}_{d-2}
\end{equation}
where the function $f(r)$ is given by

\begin{equation}
\label{eq1} f(r) = 1 - \frac{2m}{r^{d-3}} - \frac{\Lambda
r^{2}}{3} \,.
\end{equation}
In the above expression, the constant $m$ is defined by the black
hole mass M ($m = {2MG}/c^{2}$). If $\Lambda$ is positive, the
spacetime is asymptotically de Sitter in d-dimensions. In the
limit $\Lambda \rightarrow 0$ the metric (\ref{met1}) goes to the
Schwarzschild d-dimensional spacetime. In addition, if $r
\rightarrow \infty$ we have the Minkowski flat manifold in $d$
dimensions ($d \geq 4$). The quantity $d\Omega^{2}_{d-2}$ is the
standard metric for a $(d - 2)$-dimensional unit sphere:
$d\Omega^{2}_{d-2} = (d\theta^{1})^{2} +
\sin^{2}\theta^{1}(d\theta^{1})^{2} + ...+ \sin^{2}\theta^{1} +
... +\sin^{2}\theta^{d -2}d(\theta^{d -2})^{2}$. For $d = 4$ it
reduces to $d\Omega^{2} = d\theta^{2} + \sin^{2}\theta d\phi^{2}$.

\section{Schwarzschild de-Sitter Spacetime}

In this geometry, the cosmological constant is positive and can be
written as $\Lambda = 3/a^{2}$ where $a$ is the ``cosmological
radius". In order to discuss the existence of turning points we
consider the classification scheme of possible horizons for a SdS
spacetime as recently proposed by Molina\cite{m}. The number of
horizons are easily determined by the real roots of the function
$f(r) = 1 - {2m}/r^{d-3} - r^{2}/a^{2}$. Assuming $d\geq 4, a^{2}
> 0$ and $m > 0$ one has:

$\bullet$ The spacetime has two horizons if and only if the
condition

\begin{equation}
\label{pro1} \frac{m^{2}}{a^{2(d-3)}} <
\frac{(d-3)^{d-3}}{(d-1)^{d-1}}
\end{equation}
is satisfied.

$\bullet$ The spacetime has one horizon if and only if the
condition

\begin{equation}
\label{pro2} \frac{m^{2}}{a^{2(d-3)}} =
\frac{(d-3)^{d-3}}{(d-1)^{d-1}}
\end{equation}
is satisfied. This case is the extreme Schwarzschild de-Sitter
black hole.

$\bullet$ The spacetime has no horizon if and only if the
condition
\begin{equation}
\label{pro3} \frac{m^{2}}{a^{2(d-3)}} >
\frac{(d-3)^{d-3}}{(d-1)^{d-1}}
\end{equation}
is satisfied. Such spacetime has a naked singularity at $r=0$.

The roots of $f(r)$ in the above quoted cases are known as
pseudo-singularities because the metric $(\ref{met1})$ diverges
when $f=0$, although considering that all the scalars of curvature
are finite. Hence, as happens in the 4-dimensional case, the
physical singularity is located at $r=0$.

Now, since the turning points are determined by the condition
$\dot{r} = 0$, we need to consider only the radial equation of
motion\cite{c}

\begin{equation}
\label{geo1} \left(\frac{dr}{dt}\right)^{2} = E^{2} - f(r)
\end{equation}
where $E$ is an integration constant. Inserting $f(r)$ and the
value of $\Lambda$ one has
\begin{equation}
\label{geo2} \left(\frac{dr}{dt}\right)^{2} = E^{2} - 1 +
\frac{2m}{r^{d-3}} + \frac{r^{2}}{a^{2}}\,.
\end{equation}
Let us now consider the condition $\dot{r} = 0$, or equivalently,

\begin{equation}
g(r)= E^{2} - 1 + \frac{2m}{r^{d-3}} + \frac{r^{2}}{a^{2}}=0\,.
\end{equation}
Comparing the above expression with equation (\ref{eq1}), we see
that the only difference between $g(r)$ and the function $f(r)$ is
the term $E^{2}$. This means that the existence of turning points
can also be classified adopting the scheme applied to the
horizons. In this case, assuming $d\geq 4, a^{2} > 0$, $ m > 0$
and $E^{2} < 1$ the following statements are true:

$\bullet$ The function $g(r)$ has two real and positive zeros if
and only if the condition

\begin{equation}
\label{pro11} \frac{m^{2}}{a^{2(d-3)}} < \frac{(d-3)^{d-3}(1-
E^{2})^{d-1}}{(d-1)^{d-1}}
\end{equation}
is satisfied.

$\bullet$ The function $g(r)$ has one real and positive zero if
and only if the condition

\begin{equation}
\label{pro21} \frac{m^{2}}{a^{2(d-3)}} = \frac{(d-3)^{d-3}(1-
E^{2})^{d-1}}{(d-1)^{d-1}}
\end{equation}
is satisfied.

$\bullet$ The function $g(r)$ has no real and positive zero if and
only if the condition

\begin{equation}
\label{pro31} \frac{m^{2}}{a^{2(d-3)}} > \frac{(d-3)^{d-3}(1-
E^{2})^{d-1}}{(d-1)^{d-1}}
\end{equation}
is satisfied.

These real and positive roots are the turning points because the
radial component of the 3-velocity vanishes. Naturally, the case
where $E^{2} > 1$ should also be considered. However, as one may
show, in this case there are no turning points regardless of the
value of $\Lambda$ (the roots of $g(r)$ are complex or negative).

\section{Schwarzschild anti-de Sitter Spacetime}

Following the same procedure of the previews section one can study
the turning points in the SAdS spacetime. The only difference here
is that the cosmological constant must assume only negative
values, thereby leading to a change in the basic propositions.
Since the cosmological constant is negative, it is convenient to
introduce a ``pseudo radius" $b > 0$ such that $\Lambda
=-{3}/{b^{2}}$. In terms of $b$, the metric element $f(r)$ can be
written as:

\begin{equation}
f(r) = 1 - \frac{2m}{r^{d-3}} + \frac{r^{2}}{b^{2}}\,.
\end{equation}

Let us now consider the Molina\cite{tm} statement for a SAdS
spacetime: Assuming $m>0$, $d>4$ and $\Lambda < 0$, one has:

$\bullet$ If $d$ is pair or odd, the spacetime has just one
horizon.

In principle, the function $f(r)$ could have more than one real
and positive root. However, the above proposition determines the
existence of just one real and positive root which means that SAdS
spacetime has only one event horizon regardless of the number of
spatial dimensions.

On the other hand, the time-like geodesic equations for SAdS
spacetime take the same form as the one for a SdS spacetime (see
equations (\ref{geo1}) and (\ref{geo2})). The unique difference
appearing in $f(r)$ is due to the negative value of the
cosmological constant, as well as the constant $E^{2}$ which now
can assume any positive value.

In this way, we can enunciate the following statement: Assuming
$m>0$, $d>4$ and $\Lambda < 0$, one has:

$\bullet$ If $d$ is pair or odd, the function $g(r)$ has one real
and positive root. This zero of $g(r)$ defines a turning point.

\section{Conclusion}

We have discussed some aspects of the time-like geodesics for
d-dimensional SdS and SAdS spacetimes. In the first case
($\Lambda>0$) there are many possibilities. Firstly, if $E^{2}>1$
there are no turning points. But if $E^{2}<1$, the number of
turning points is equal to the number of horizons. For SAdS in d
dimensions, the black hole manifold has only one horizon and one
turning point. This result holds regardless of the value of
$E^{2}$. Therefore, for all these cases, the number of possible
turning points (when they exist!) is just the same number of
horizons.

It should be also interesting to extend these results for a
d-dimensional charged black hole with $\Lambda$. We recall that in
the Reissner-Nordst\"{o}m spacetime (4-dimensions) there is a
turning point inside the internal radius (see, for instance,
\cite{c} and \cite{b}). In general, one expects a number of
horizons different from the number of turning points.

{\bf Acknowledgements:}This work was supported by CAPES and CNPq
(Brazilian Research Agencies).

\end{document}